\documentclass[11pt,a4paper]{article}
\usepackage{amssymb}
\usepackage{amsfonts}
\usepackage{amsmath}
\usepackage{amsfonts}
\usepackage{amsmath,amssymb}
\usepackage{epsfig,graphicx}

\input{tcilatex}

\begin{document}

\begin{center}
{\LARGE On gravity unification in SL(2N,C)}\ {\LARGE gauge theories}

{\huge \ }

{\Huge \bigskip }

\bigskip

\bigskip

\textbf{J.L.~Chkareuli}$^{1,2}$

\bigskip

$^{1}$\textit{Institute of Theoretical\ Physics, Ilia State University, 0162
Tbilisi, Georgia}

$^{2}$\textit{Andronikashvili} \textit{Institute of Physics, Tbilisi State
University, 0177 Tbilisi, Georgia\ \ }

\bigskip {\scriptsize \bigskip }\bigskip \bigskip \bigskip \bigskip \bigskip

\textbf{Abstract}
\end{center}

\bigskip

The local $SL(2N,C)$ symmetry is shown to provide, when appropriately
constrained, a viable framework for a consistent unification of the known
elementary forces, including gravity. Such a covariant constraint implies
that an actual gauge field multiplet in the $SL(2N,C)$ theory is ultimately
determined by the associated tetrad fields which not only specify the
geometric features of spacetime but also govern which local internal
symmetries are permissible within it. As a consequence, upon the covariant
removal of all "redundant" gauge field components, the entire theory only
exhibits the effective $SL(2,C)\times SU(N)$ symmetry, comprising $SL(2,C)$
gauge gravity on one hand and $SU(N)$ grand unified theory on the other.
Given that all states involved in the $SL(2N,C)$ theories are additionally
classified according to their spin values, many potential $SU(N)$ GUTs,
including the conventional $SU(5)$ theory, appear to be irrelevant for
standard spin $1/2$ quarks and leptons. Meanwhile, applying the $SL(2N,C)$
symmetry to the model of composite quarks and leptons with constituent
chiral preons in its fundamental representations reveals, under certain
natural conditions, that among all accompanying $SU(N)_{L}\times SU(N)_{R}$
chiral symmetries of preons and their composites only the $SU(8)_{L}\times
SU(8)_{R}$ meets the anomaly matching condition ensuring masslessness of
these composites at large distances. This, in turn, identifies $SL(16,C)$
with the effective $SL(2,C)\times SU(8)$ symmetry, accommodating all three
quark-lepton families, as the most likely candidate for hyperunification of
the existing elementary forces.

%
%%%%%%%%%%%%%%%%%%%%%%%%%%%%%%%%%%%%%%%%%%%%%%%%%
%%%%%%
\thispagestyle{empty}\newpage

\section{Introduction}

As is well known, there exists a certain similarity between gravity and
three other elementary forces when considering gravity within a conventional
gauge theory framework \cite{Utiyama, Kibble, ish}. Indeed, the
spin-connection fields gauging the local $SL(2,C)$ symmetry group of gravity
emerge much like photons and gluons in the Standard Model. It is, therefore,
conceivable that these spin-connections could be unified with the ordinary
SM gauge bosons in a certain non-compact symmetry group, thereby leading to
the hyperunification of all known elementary gauge forces. In the following,
we refer to such theories as hyperunified theories (HUTs), and specifically
as the $SL(2N,C)$ HUT when speaking about integration of the $SL(2,C)$ gauge
gravity with the $SU(N)$ grand unified theory, respectively. We also
designate $SU(N)$ as the "hyperflavor" symmetry and fields located in its
representations as the "hyperflavored" fields.

In fact, there are many classes of models in the literature where
unification of gravity and other interactions goes through a unification of
the local Lorentz and internal symmetries in the framework of some
non-compact covering symmetry group \cite{cho, hu, per, cham}. Their
difficulties are well known and, to varying degrees, they generally appear
in the $SL(2N,C)$ HUT as well \cite{jpl}. Firstly, the vector fields in the
total gauge multiplet of this group are always accompanied by the
axial-vector fields which must be somehow excluded from the theory as there
is no direct indication whatsoever of their existence. Then, while vector
fields are proposed to mediate ordinary gauge interactions, tensor fields
must provide the minute gravity interactions to align with reality. The
crucial point lies in the fact that, whereas in pure gravity case, one can
solely consider the action being linear in the curvature ($R$) constructed
from the tensor field, the unification with other interactions necessitates
the inclusion of quadratic curvature ($R^{2}$) terms as well. Consequently,
tensor fields in these terms will induce interactions comparable to those of
the gauge vector fields in the Standard Model. Moreover, the tensor fields,
akin to the vector ones, exhibit now the internal $SU(N)$ symmetry features
implying the existence of the multiplet of hyperflavored gravitons rather
than a single neutral one. Apart from that, such $R+R^{2}$ Lagrangians for
gravity are generally known to contain ghosts and tachyons rendering them
essentially unstable. And lastly, but perhaps most importantly, a potential
pitfall in hyperunified theories stems from the Coleman-Mandula theorem \cite%
{18} concerning the impossibility of merging spacetime and internal
symmetries. It is worth noting that this theorem initially surfaced
precisely in connection with one of the special cases of $SL(2N,C)$,
specifically the $SL(6,C)$ symmetry \cite{ish1}, used a long time ago as a
possible relativistic version of the global $SU(6)$ symmetry model
describing the spin-unitary spin symmetry classification of mesons and
baryons \cite{feza}.

In contrast, we aim to demonstrate here how, in the $SL(2N,C)$ HUT
framework, these difficulties can be naturally overcome in the way as yet
unexplored. The key idea is that the extended gauge multiplet in the $%
SL(2N,C)$ theory -- comprising generally the vector, axial-vector and tensor
field submultiplets -- is suitably constrained by the associated tetrad
multiplets which are assumed to not only determine the geometric features of
spacetime, but also control which local internal symmetries and associated
gauge field interactions are permitted in it. Specifically, we propose that
the actual gauge multiplet $I_{\mu }$ arises as result of the tetrad
filtering of some "prototype" nondynamical multiplet $\mathcal{I}_{\mu }$
which globally transforms similarly to $I_{\mu }$, but, unlike it, does not
itself gauge the corresponding fermion system. These two multiplets are
connected in a covariant way using the tetrads $e_{\sigma }$ and $e^{\sigma
} $ that, instead of being imposed by postulate, can be incorporated into
the theory through the Lagrange multiplier type term 
\begin{equation}
\mathrm{C}\left( I_{\mu }-\frac{1}{4}e_{\sigma }\mathcal{I}_{\mu }e^{\sigma
}\right) ^{2}\text{\ \ \ }  \label{is}
\end{equation}%
where $\mathrm{C}$ is an arbitrary constant. This term, upon variation under
the multiplet $\mathcal{I}_{\mu }$, yields the filtering condition, $I_{\mu
}=e_{\sigma }\mathcal{I}_{\mu }e^{\sigma }/4$, mentioned above.
Consequently, the gauge multiplet $I_{\mu }$ retains only those components
of the prototype multiplet $\mathcal{I}_{\mu }$ which result from the tetrad
filtering. In other words, the prototype multiplet $\mathcal{I}_{\mu }$
becomes partially dynamical to the extent permitted by the tetrads.

Now, in the case of standard or strictly invertible tetrads, such filtering
excludes the gravity induced tensor fields in the gauge multiplet $I_{\mu }$%
, as we demonstrate below. However, when the tetrad invertibility condition
is slightly broken, the appropriately weakened tensor fields come into play.
This occurs in a way that their interaction essentially decouples from other
elementary forces and effectively adheres to the Einstein-Cartan type
gravity action. The corresponding curvature-squared terms constructed from
the filtered tensor fields appear to be vanishingly small and can be
disregarded compared to the standard strength-squared terms for vector
fields. As a result, the entire theory effectively exhibits a local $%
SL(2,C)\times SU(N)$ symmetry rather than the unified $SL(2N,C)$ symmetry,
which solely provides the structure of total multiplets for gauge and matter
fields in the theory. Consequently, this naturally leads to $SL(2,C)$ gauge
gravity on one hand and $SU(N)$ grand unified theory (GUT) on the other,
thereby effectively bypassing the constraints of the Coleman-Mandula theorem.

The paper is organized as follows. In Section 2 we provide a standard
presentation of the $SL(2,C)$ gauge gravity which is then discussed using
the filtering approach. Section 3 introduces the filtered $SL(2N,C)$ HUT
which ultimately lead to the $SL(2,C)$ gauge gravity in conjunction with the 
$SU(N)$ grand unified theory. Section 4 examines specific HUT models with
particular focus on the $SL(16,C)$ theory giving rise to the $SU(8)$ GUT
with all three families of composite quarks and leptons that is studied in
some detail. Finally, our summary is presented in Section 5.

\section{$SL(2,C)$ gravity}

\subsection{Standard framework}

We first present the $SL(2,C)$ gravity model, partially following the
pioneering work \cite{ish}. Let there be a local frame at any spacetime
point where the global $SL(2,C)$ symmetry group acts. According to this
symmetry, the basic fermions of the theory transform as 
\begin{equation}
\Psi \rightarrow \Omega \Psi \text{, \ \ }\Omega =\exp \left\{ \frac{i}{4}%
\theta _{ab}\gamma ^{ab}\right\} \text{ }  \label{om}
\end{equation}%
\newline
where the matrix $\Omega $ satisfies a pseudounitarity condition, $\Omega
^{-1}=\gamma _{0}\Omega ^{+}\gamma _{0}$ (the transformation parameters\ $%
\theta _{ab}$ are assumed to be constant for now). Furthermore, to ensure
the invariance of their kinetic terms, $i\overline{\Psi }\gamma ^{\mu
}\partial _{\mu }\Psi $, one needs to replace $\gamma $-matrices in them
with a set of some tetrad matrices $e^{\mu \text{ }}$which transform like%
\begin{equation}
e^{\mu }\rightarrow \Omega e^{\mu }\Omega ^{-1}\text{ }  \label{trl}
\end{equation}%
Generally, the tetrad matrices $e^{\mu \text{ }}$, as well as their
conjugates $e_{\mu }^{\text{ }}$, contain the appropriate tetrad fields $%
e_{a}^{\mu }$ and $e_{\mu }^{a}$, respectively,%
\begin{equation}
e_{\mu }=e_{\mu }^{a}\gamma _{a}\text{ , \ }e^{\mu }=e_{a}^{\mu }\gamma ^{a}%
\text{ }  \label{tlf}
\end{equation}%
which transforms infinitesimally as 
\begin{equation}
\delta e^{\mu c}=\frac{1}{2}\theta _{ab}(e^{\mu a}\eta ^{bc}-e^{\mu b}\eta
^{ac})  \label{tr1a}
\end{equation}%
They, as usual, satisfy the invertibility conditions%
\begin{equation}
e_{\mu }^{a}e_{a}^{\nu }=\delta _{\mu }^{\nu }\text{, \ }e_{\mu
}^{a}e_{b}^{\mu }=\delta _{b}^{a}  \label{or}
\end{equation}%
and determine the metric tensors in the theory 
\begin{equation}
g_{\mu \nu }=\frac{1}{4}Tr(e_{\mu }e_{\nu })=e_{\mu }^{a}e_{\nu }^{b}\eta
_{ab}\text{\ , \ }g^{\mu \nu }=\frac{1}{4}Tr(e^{\mu }e^{\nu })=e_{a}^{\mu
}e_{b}^{\nu }\eta ^{ab}  \label{gmn}
\end{equation}

Going now to the case when the $SL(2,C)$ transformations (\ref{om}) become
local, $\theta _{ab}\equiv \theta _{ab}(x)$, one have to introduce the
spin-connection gauge field multiplet $I_{\mu }$\ transforming as usual%
\begin{equation}
I_{\mu }\rightarrow \Omega I_{\mu }\Omega ^{-1}-\frac{1}{ig}(\partial _{\mu
}\Omega )\Omega ^{-1}  \label{123}
\end{equation}%
thus providing the fermion field by covariant derivative 
\begin{equation}
\partial _{\mu }\Psi \rightarrow D_{\mu }\Psi =\partial _{\mu }\Psi
+igI_{\mu }\Psi \text{ }  \label{124}
\end{equation}%
where $g$ presents the gauge coupling constant extracted for later
convenience. The $I_{\mu }$ multiplet gauging the $SL(2,C)$ has by
definition the form 
\begin{equation}
I_{\mu }=\frac{1}{4}T_{\mu \lbrack ab]}\gamma ^{ab}\text{ \ \ }  \label{125}
\end{equation}%
where the flat spacetime tensor field components $T_{\mu \lbrack ab]}$
transform as%
\begin{equation}
\delta T_{\mu }^{[ab]}=\frac{1}{2}\theta _{\lbrack cd]}[(T_{\mu }^{[ac]}\eta
^{bd}-T_{\mu }^{[ad]}\eta ^{bc})-(T_{\mu }^{[bc]}\eta ^{ad}-T_{\mu
}^{[bd]}\eta ^{ac})]-\frac{1}{g}\partial _{\mu }\theta ^{\lbrack ab]}
\label{123a}
\end{equation}

The tensor field\ $T_{\mu \lbrack ab]}$ may in principle propagate, while
the tetrad $e^{\mu }$ is not considered as a dynamical field. So, the
invariant Lagrangian built from its strength 
\begin{eqnarray}
I_{\mu \nu } &=&\partial _{\lbrack \mu }I_{\nu ]}+ig[I_{\mu },I_{\nu }]=%
\frac{1}{4}T_{\mu \nu }^{[ab]}\gamma _{ab}  \notag \\
T_{\mu \nu }^{[ab]} &=&\partial _{\lbrack \nu }T_{\mu ]}^{[ab]}+g\eta
_{cd}T_{[\mu }^{[ac]}T_{\nu ]}^{[bd]}  \label{137}
\end{eqnarray}%
can be written in a conventional form%
\begin{equation}
e\mathcal{L}_{G}=\frac{1}{2\kappa }e_{[a}^{\mu }e_{b]}^{\nu }\text{ }T_{\mu
\nu }^{[ab]}\text{ , \ \ }e\equiv \lbrack -\det Tr(e^{\mu }e^{\nu
})/4]^{-1/2}\text{ }  \label{127}
\end{equation}%
(where $\kappa $\ stands for the modified Newtonian constant $8\pi
/M_{Pl}^{2}$) once the commutator for tetrads and some of standard relations
for $\gamma $-matrices have been used\footnote{%
We give here some of them used throughout the paper%
\begin{eqnarray*}
\gamma ^{ab} &=&i[\gamma ^{a},\gamma ^{b}]/2,\text{ }\gamma ^{a}\gamma
^{b}=\gamma ^{ab}/i+\eta ^{ab}\widehat{1},\text{ }\gamma _{c}\gamma
^{\lbrack ab]}\gamma ^{c}=0 \\
\lbrack \gamma ^{ab},\gamma ^{a^{\prime }b^{\prime }}] &=&2i(\eta
^{ab^{\prime }}\gamma ^{ba^{\prime }}+\eta ^{ba^{\prime }}\gamma
^{ab^{\prime }}-\eta ^{aa^{\prime }}\gamma ^{bb^{\prime }}-\eta ^{bb^{\prime
}}\gamma ^{aa^{\prime }}) \\
Tr(\gamma ^{ab}\gamma ^{a^{\prime }b^{\prime }}) &=&4(\eta ^{aa^{\prime
}}\eta ^{bb^{\prime }}-\eta ^{ab^{\prime }}\eta ^{ba^{\prime }}),\text{ \ }%
Tr(\gamma ^{ab}\gamma _{cd})=4(\delta _{c}^{a}\delta _{d}^{b}-\delta
_{c}^{b}\delta _{d}^{a}) \\
Tr(\gamma ^{ab}\gamma ^{a^{\prime }b^{\prime }}\gamma ^{a^{\prime \prime
}b^{\prime \prime }}) &=&4i[\eta ^{aa^{\prime }}(\eta ^{a^{\prime \prime
}b^{\prime }}\eta ^{bb^{\prime \prime }}-\eta ^{a^{\prime \prime }b}\eta
^{b^{\prime }b^{\prime \prime }})+\eta ^{ab^{\prime }}(\eta ^{a^{\prime
\prime }b}\eta ^{a^{\prime }b^{\prime \prime }}-\eta ^{a^{\prime }a^{\prime
\prime }}\eta ^{bb^{\prime \prime }}) \\
&&+\eta ^{a^{\prime }b}(\eta ^{aa^{\prime \prime }}\eta ^{b^{\prime
}b^{\prime \prime }}-\eta ^{a^{\prime \prime }b^{\prime }}\eta ^{ab^{\prime
\prime }})+\eta ^{bb^{\prime }}(\eta ^{a^{\prime }a^{\prime \prime }}\eta
^{ab^{\prime \prime }}-\eta ^{a^{\prime }b^{\prime \prime }}\eta
^{aa^{\prime \prime }})]
\end{eqnarray*}%
where $\widehat{1}$ in the above is the $4\times 4$ unit matrix.}. This is,
in fact, the simplest pure gravity Lagrangian taken in the Palatini type
formulation. Indeed, its variation under the tensor field gives the
constraint allowing to express it through the tetrad and its derivative that
reduces $e\mathcal{L}_{G}$ to the standard Einstein Lagrangian. It is
written with the scalar\ factor $e$ which, while irrelevant for $SL(2,C)$
gauge invariance itself, provides an extra invariance of the action under
general four-coordinate transformations $GL(4,R)$ as well \cite{ish}.

Meanwhile, in presence of fermions, the gauge invariant fermion matter
coupling given by the covariant derivative (\ref{124}, \ref{125}) implies
the extra tensor field interaction with\ the spin-current density 
\begin{equation}
e\mathcal{L}_{M}^{int}=-\frac{1}{2}g\epsilon ^{abcd}T_{\mu \lbrack
ab]}e_{c}^{\mu }\overline{\Psi }\gamma _{d}\gamma _{5}\Psi  \label{138}
\end{equation}%
This is a key feature of the Einstein-Cartan type gravity \cite{Kibble}
which eventually results in, apart from the standard GR, the tiny
four-fermion (spin current-current) interaction in the matter sector 
\begin{equation}
\kappa \left( \overline{\Psi }\gamma _{d}\gamma ^{5}\Psi \right) (\overline{%
\Psi }\gamma ^{d}\gamma ^{5}\Psi )  \label{4f}
\end{equation}

On the other hand, one could augment the linear curvature Lagrangian (\ref%
{127}) with the curvature squared terms, thereby rendering some tensor field
components dynamical. It is known that theories of this type are strongly
constrained to a specific form to ensure freedom from ghosts and tachyons 
\cite{nev}. Expressed in terms of tensor field strengths, the acceptable
quadratic curvature part in them appears as 
\begin{equation}
e\mathcal{L}^{(2)}=Q(T_{ab}^{\mu \nu }T_{\mu \nu }^{ab}+T_{ab}^{\mu \nu
}T_{\rho \sigma }^{cd}e_{\mu c}e_{\nu d}e^{\rho a}e^{\sigma b}-4T_{ab}^{\mu
\nu }T_{\rho \nu }^{ac}e_{\mu c}e^{\rho b})  \label{re}
\end{equation}%
Actually, such a theory contains, apart the graviton, some superheavy scalar
excitation $S(0^{-})$ with a mass, $m_{S}^{2}\sim M_{P}^{2}/Q$, that can
hardly be observed unless the numerical parameter $Q$ in (\ref{re}) is
exceedingly large.

\subsection{ Tetrad filtering case}

As previously mentioned, the $SL(2,C)$ gauge gravity is implied as a part of
the unified set of all elementary forces assembled in the $SL(2N,C)$
symmetry framework. This raises a key issue of how to reconcile the
exceptionally weak gravitational force related to the tensor fields with the
significant Standard Model forces associated with the vector field
submultiplet. The corresponding Lagrangian generally includes terms that are
linear and quadratic in the strength of the entire gauge field multiplet
(each with independent coupling constants). Interestingly, as is shown
later, the linear terms pose no problem, as they only generate gauge
gravity, governed by its own coupling constant (related to the Planck mass,
as usual). However, the quadratic terms for the tensor fields share the same
dimensionless coupling constant as the vector field interactions, which is
certainly unacceptable. We propose that the tensor field quadratic terms
might become negligible due to the minuscule nature of the filtered tensor
fields themselves, while the vector fields remain unaffected by this
filtering process. Whereas it would be particularly noteworthy to
immediately consider this scenario within the entire framework of $SL(2N,C)$
theory, we nonetheless begin by examining the pure tensor field case to
elucidate the proposed tetrad filtering mechanism in greater detail.

In this regard, we propose that the gauge field multiplet $I_{\mu }$ (\ref%
{125}) stems from some prototype nondynamical multiplet $\mathcal{I}_{\mu }$%
, which globally transforms akin to $I_{\mu }$, but does not gauge the
matter fermions. They are connected in a covariant way through the tetrads
involved 
\begin{equation}
I_{\mu }=\frac{1}{4}e_{\sigma }\mathcal{I}_{\mu }e^{\sigma },\text{ }%
\mathcal{I}_{\mu }=\frac{1}{4}\mathcal{T}_{\mu \lbrack ab]}\gamma ^{ab}
\label{u}
\end{equation}%
when the Lagrange multiplier term (\ref{is}) in the gauge gravity Lagrangian
is varied under the nondynamical $\mathcal{I}_{\mu }$ multiplet. However,
one can easily find that, as follows from the $\gamma $-matrix algebra$^{1}$%
, and, in particular, from the identity 
\begin{equation}
\gamma _{c}\gamma ^{\lbrack ab]}\gamma ^{c}=0  \label{gmm}
\end{equation}%
the gauge multiplet $I_{\mu }$ with the strictly invertible tetrads (\ref{or}%
) automatically vanishes. This means that the theory only possesses the
global $SL(2,C)$ symmetry in this limit and gauge gravity is in fact absent.

The gauge multiplet $I_{\mu }$ may only appears if the tetrads are no longer
orthonormal but their invertibility conditions (\ref{or}) include some
vanishingly small deviations 
\begin{equation}
e_{\mu }^{a}e_{b}^{\mu }=\delta _{b}^{a}+\varepsilon q_{b}^{a},\text{ \ }%
e_{\mu }^{a}e_{a}^{\nu }=\delta _{\mu }^{\nu }+\varepsilon p_{\mu }^{\nu }%
\text{ \ \ \ \ }(\varepsilon \ll 1)  \label{eem}
\end{equation}%
given by the infinitesimal tensors $\varepsilon q_{b}^{a}$ and $\varepsilon
p_{\mu }^{\nu }$ (with the tiny constant parameter $\varepsilon $) ,
respectively. Indeed, now the gauge multiplet $I_{\mu }$ comes to 
\begin{equation}
I_{\mu }=\frac{1}{4}T_{\mu \lbrack ab]}\gamma ^{ab}=\frac{\varepsilon }{16}%
\mathcal{T}_{\mu \lbrack ab]}q_{d}^{c}(\gamma _{c}\gamma ^{ab}\gamma ^{d})%
\text{\ }  \label{b}
\end{equation}%
being solely determined by the tiny tensor $\varepsilon q_{b}^{a}$.
Multiplying the both sides by $\gamma ^{a^{\prime }b^{\prime }}$ and taking
the traces in them one can readily find using the $\gamma $ matrix algebra$%
^{2}$ the relation between the tensor fields themselves 
\begin{subequations}
\begin{equation}
T_{\mu \lbrack ab]}=\frac{\varepsilon }{2}(\mathcal{T}_{\mu \lbrack
bc]}q_{a}^{c}-\mathcal{T}_{\mu \lbrack ac]}q_{b}^{c}+\mathcal{T}_{\mu
\lbrack ab]}q_{c}^{c}/2)=\varepsilon \mathbf{T}_{\mu \lbrack ab]}  \label{bb}
\end{equation}%
Thus, the starting prototype, while nondynamical, tensor field multiplet $%
\mathcal{T}_{\mu \lbrack ab]}$ is predominantly extinguished and only its
minuscule portion emerges in the gauge multiplet $T_{\mu \lbrack ab]}$.
Remarkably, the filtering process triggers an important suppression
mechanism in the $SL(2,C)$ gauge gravity theory that allows it to be
reformulated solely in terms of the weakened tensor field multiplet $%
\varepsilon \mathbf{T}_{\mu \lbrack ab]}$. This multiplet is, in fact, the
product of the prototype tensor field $\mathcal{T}_{\mu \lbrack ab]}$ with
the infinitesimal tensor $\varepsilon q_{b}^{a}$ which describes the tiny
deviation in the modified invertibility conditions (\ref{eem}) for tetrads.\ 

Analogously, the metric tensor will also include such a deviation which we
define from the similar equation 
\end{subequations}
\begin{equation}
e_{\mu }^{a}e_{a\nu }=g_{\mu \nu }+r_{\mu \nu },\text{ }e_{a}^{\mu }e^{a\nu
}=g^{\mu \nu }+r^{\mu \nu }  \label{eem1}
\end{equation}%
Multiplying the basic equations (\ref{eem}) by the proper tetrads one can
readily find relations between the deviations 
\begin{equation}
p_{\mu }^{\nu }e_{\nu }^{a}=q_{b}^{a}e_{\mu }^{b},\text{ \ }(pp)_{\mu }^{\nu
}e_{\nu }^{a}=(qq)_{b}^{a}e_{\mu }^{b},\text{ \ }(p...p)_{\mu }^{\nu }e_{\nu
}^{a}=(q...q)_{b}^{a}e_{\mu }^{b}  \label{rr}
\end{equation}%
so that all deviations can be expressed in terms of the $q$ parameter only.
Such $q$-depending deviation appears in the metric tensor in (\ref{eem1}) as
well provided that one requires general covariance for the metric tensor $%
g_{\mu \nu }$

\begin{equation}
g_{\mu \nu }e^{\nu b}=e_{\mu }^{b},\text{ }r_{\mu \nu }e^{\nu b}=-e_{\mu
}^{b}+e_{\mu }^{a}e_{a\nu }e^{\nu b}=\varepsilon e_{\mu }^{a}q_{a}^{b}
\label{rr1}
\end{equation}%
where we have also used the deviation equations (\ref{eem}). Multiplying
then the both sides by $e_{b\rho }$ one finds%
\begin{equation}
r_{\mu \nu }(\delta _{\rho }^{\nu }+p_{\rho }^{\nu })=e_{\mu
}^{a}q_{a}^{b}e_{b\rho }  \label{r1a}
\end{equation}%
which after multiplying by the conjugated factor $\delta _{\sigma }^{\rho
}-p_{\sigma }^{\rho }$ finally gives 
\begin{equation}
r_{\mu \nu }\simeq \lbrack \varepsilon q_{a}^{b}-\varepsilon
^{2}(qq)_{a}^{b}]e_{\mu }^{a}e_{b\nu }  \label{r1b}
\end{equation}%
up to the second order terms in $q$. With the same accuracy the metric
tensors acquires the form \ 
\begin{equation}
g_{\mu \nu }\simeq e_{\mu }^{a}e_{\nu }^{c}\eta _{bc}[\delta
_{a}^{b}-\varepsilon q_{a}^{b}+\varepsilon ^{2}(qq)_{a}^{b}],\text{ \ }%
g_{\mu \nu }g^{\nu \rho }=\delta _{\mu }^{\rho }+O(\varepsilon ^{2})
\label{rr22}
\end{equation}%
and, respectively, 
\begin{equation}
\eta _{ab}\simeq g_{\mu \nu }e_{a}^{\mu }e_{c}^{\nu }[\delta
_{b}^{c}-\varepsilon q_{b}^{c}+\varepsilon ^{2}(qq)_{b}^{c}],\text{ }\eta
_{ab}\eta ^{bc}=\delta _{a}^{c}+O(\varepsilon ^{2})  \label{rr22'}
\end{equation}%
The exact expressions for them can be symbolically written as 
\begin{equation}
g_{\mu \nu }=e_{\mu }^{a}e_{\nu }^{c}\eta _{bc}\left[ \frac{1}{1+\varepsilon
q}\right] _{a}^{b},\text{ \ }\eta _{ab}=g_{\mu \nu }e_{a}^{\mu }e_{c}^{\nu }%
\left[ \frac{1}{1+\varepsilon q}\right] _{b}^{c}  \label{rr4}
\end{equation}%
transparently showing their modification compared to the standard case (\ref%
{gmn}).

In the following, we assume for simplicity that the above deviation tensors
are traceless ($q_{a}^{a}=p_{\mu }^{\mu }=0$) being arisen from some
symmetric traceless tensors%
\begin{equation}
q_{b}^{a}=\eta _{bc}q^{\{ac\}},\text{ }p_{\mu }^{\nu }=g_{\mu \sigma
}p^{\{\nu \sigma \}}  \label{q'}
\end{equation}%
For such a choice, despite the nonzero deviations in (\ref{eem}) the tetrad
matrix invertibility condition remains its form 
\begin{equation}
\frac{1}{4}e_{\mu }e^{\mu }=\widehat{1}+\varepsilon q_{b}^{a}\gamma
_{a}\gamma ^{b}/4=\widehat{1}\text{ \ }  \label{q}
\end{equation}

Interestingly, the modified invertibility conditions (\ref{eem}) may be
considered as those that appear due to condensation of the tetrad which can
be written in the form 
\begin{subequations}
\begin{equation}
\text{\ }e_{\mu }^{a}\,=\delta _{\mu }^{a}h+\widehat{e}_{\mu }^{a}\,,\ \text{%
\ }\delta _{a}^{\mu }\widehat{e}_{\mu }^{a}=0  \label{bd}
\end{equation}%
where 
\end{subequations}
\begin{equation}
h=(1-\widehat{e}_{\mu }^{a}\widehat{e}_{a}^{\mu }/4)^{1/2}  \label{h1}
\end{equation}%
Such a form provides the vacuum expectation value for the tetrad field $%
e_{\mu }$ 
\begin{equation}
\left\langle e_{\mu }\right\rangle =\gamma _{\mu }  \label{vev}
\end{equation}%
which represents an extremum of the action with $\widehat{e}_{\mu }^{a}$ and 
$h$ being as the effective zero mode and Higgs mode, respectively. This
could be considered quite an ordinary example, if tetrads were treated as
the dynamical fields in the $SL(2,C)$ gauge theory that is not actually
supposed. However, nothing prevents the above conditions from being seen as
a would-be spontaneous breakdown of the local frame $SL(2,C)$ symmetry for
tetrads, while the theory is still left Poincare-invariant.

Although tetrads are not considered as dynamical fields in the theory, one
can take, nonetheless, that the above conditions cause a would-be
spontaneous breakdown of the local frame $SL(2N,C)$ symmetry for tetrads.

The basic Lagrangian for the $SL(2,C)$ gravity will now result in the
appropriate analogs of the gravity and matter field Lagrangians (\ref{127}, %
\ref{138}), respectively. Indeed, the minimal gravity Lagrangian (\ref{127})
remains practically the same form 
\begin{equation}
e\mathcal{L}_{G}=\frac{1}{2\kappa }e_{[a}^{\mu }e_{b]}^{\nu }\mathbf{T}_{\mu
\nu }^{[ab]}  \label{12}
\end{equation}%
though the tensor field strength $T_{\mu \nu \lbrack ab]}$ has been properly
modified according to the relations (\ref{b}, \ref{bb}) taken%
\begin{equation}
T_{\mu \nu }^{[ab]}=\varepsilon \mathbf{T}_{\mu \nu }^{[ab]},\text{ \ }%
\mathbf{T}_{\mu \nu }^{[ab]}=\partial _{\lbrack \nu }\mathbf{T}_{\mu
]}^{[ab]}+\varepsilon g\eta _{cd}\mathbf{T}_{[\mu }^{[ac]}\mathbf{T}_{\nu
]}^{[bd]}  \label{127'}
\end{equation}%
Remarkably, once the tiny parameter $\varepsilon $ is absorbed in the
gravity constant\ $\kappa $, the tensor field strength $\mathbf{T}_{\mu \nu
}^{[ab]}$ acquires the modified gauge coupling constant $\varepsilon g$ for
the new tensor field $\mathbf{T}_{\mu }^{[ab]}$. Such a modification will
also appear for the matter coupling 
\begin{equation}
e\mathcal{L}_{M}^{int}=-\frac{\varepsilon g}{2}\epsilon ^{abcd}\mathbf{T}%
_{\mu \lbrack ab]}e_{c}^{\mu }\overline{\Psi }\gamma _{d}\gamma _{5}\Psi 
\label{138'}
\end{equation}%
and for the possible quadratic tensor field strength terms 
\begin{equation}
e\mathcal{L}_{T}=\varepsilon ^{2}Q_{n}\mathcal{L}_{n}^{(2)}  \label{139}
\end{equation}%
where, for generality, the latter is taken to contain all possible
combinations of the tensor field strength bilinears with tetrads, each with
an arbitrary $Q_{n}$ constant.

In this context, the propagation of the tensor field $\mathbf{T}_{\mu
}^{[ab]}$ in the hyperunified $SL(2N,C)$ theory seems to be irrelevant. In
fact, its kinetic term contained in (\ref{139}), in sharp contrast to the
ordinary kinetic terms of the vector (and axial-vector) fields, scales as $%
\varepsilon ^{2}$ and therefore can be neglected. As in the standard case
considered above, the variation of the total linear Lagrangian in (\ref{12}, %
\ref{138'}) under prototype tensor field $\mathbf{T}_{\mu }^{[ab]}$ just
leads to the constraint equation rather than the normal equation of motion.
Meanwhile, the variation of total gravity Lagrangian with respect to the
tetrad $e_{a}^{\mu }$ leads immediately to the equation of motion of the
Einstein-Cartan type gravity.

In conclusion, it is worth noting, that one might believe that all the above
modifications are merely fictitious as a simple rescaling of the tensor
field $\mathbf{T}_{\mu }=\mathbf{T}_{\mu }^{\prime }/\varepsilon $ seemingly
restores the standard case albeit with slightly broken general covariance
due to the deviations in tetrads (\ref{eem}). However, this rescaling trick
only works for pure gauge gravity with local $SL(2,C)$ symmetry and appears
inappropriate in the framework of $SL(2N,C)$ gauge theory. In reality, the
weakness of the filtered tensor field is exclusive to the tensor field
submultiplet and does not extend to the entire gauge multiplet of $SL(2N,C)$
containing also the vector and axial-vector fields. Consequently, it cannot
be circumvented by mere rescaling, as we show in more detail later in
Section 3.4.

\section{Toward $SL(2N,C)$ hyperunification}

\subsection{Basics of $SL(2N,C)$}

In general, the $SL(2N,C)$ symmetry group encompasses, among its primary
subgroups, the aforementioned $SL(2,C)$ symmetry, which covers the
orthochronous Lorentz group, and the internal $U(N)$ symmetry group
(including the hyperflavor $SU(N)$ symmetry). Indeed, the $8N^{2}-2$
generators of $SL(2N,C)$ are formed from the tensor products of the
generators of $SL(2,C)$ and generators of $U(N)$ so that the basic
transformation applied to the fermions looks as follows 
\begin{equation}
\Omega =\exp \left\{ \frac{i}{2}\left[ \left( \theta ^{k}+i\theta
_{5}^{k}\gamma _{5}\right) \lambda ^{k}+\frac{1}{2}\theta _{ab}^{K}\gamma
^{ab}\lambda ^{K}\right] \right\} \text{ \ \ }(K=0,k)  \label{rt}
\end{equation}%
Here, among the $\lambda ^{K}$ matrices, $\lambda ^{k}$ $(k=1,...,N^{2}-1$)
represent the $SU(N)$\ Gell-Mann matrices, while $\lambda ^{0}$ is the unit
matrix $\widehat{1}$ corresponding to the $U(1)$ generator (all $\theta $
parameters may be constant or, in general, depend on the spacetime
coordinate). Hereafter, we use the uppercase Latin letters ($I,J,K$) for the 
$U(N)$ symmetry case, while the lowercase letters ($i,j,k$) for the $SU(N)$
symmetry one\footnote{%
Some relations for $\lambda $ matrices used below are given here%
\begin{eqnarray*}
\lbrack \lambda ^{k},\lambda ^{l}] &=&2if^{klm}\lambda ^{m},\text{ }%
\{\lambda ^{k},\lambda ^{l}\}=2(\delta ^{kl}\widehat{1}+d^{klm}\lambda ^{m})
\\
\lambda ^{k}\lambda ^{l}\lambda ^{k} &=&-\lambda ^{l},\text{ \ }\lambda
^{K}\lambda ^{l}\lambda ^{K}=0,\text{ \ }Tr(\lambda ^{k}\lambda
^{l})=N\delta ^{kl}
\end{eqnarray*}%
The connections with a standard choice of the $SU(N)$ matrices are given by
the links 
\begin{equation*}
\lambda ^{K}=\sqrt{2N}T^{K},\text{ }f^{ijk}=\sqrt{N/2}F^{ijk},\text{ \ }%
d^{ijk}=\sqrt{N/2}D^{ijk}
\end{equation*}%
}.

For description of the fermion matter in the theory one needs again to
introduce the generalized tetrad multiplet 
\begin{equation}
e_{\mu }=(e_{\mu }^{aK}\gamma _{a}+e_{\mu 5}^{aK}\gamma _{a}\gamma
_{5})\lambda ^{K}\text{ \ }  \label{lllll}
\end{equation}%
which transforms, as before, according to (\ref{trl}) where the
transformation matrix is now given by equation (\ref{rt}). Despite its
somewhat cumbersome extension which generally appears in the $SL(2N,C)$
framework, it would be natural for tetrad flat space components in (\ref%
{lllll}) to essentially have the same form as in the pure gravity case. This
implies that such an extension might not include the axial-vector part that
could be reached through the gauge invariant constraints put on tetrads. In
fact, one can introduce for that some\ special nondynamical $SL(2N,C)$
scalar multiplet in the theory 
\begin{equation}
S=\exp \{i[(s^{k}+ip^{k}\gamma _{5})\lambda ^{k}+t_{ab}^{K}\gamma
^{ab}\lambda ^{K}/2]\}  \label{ss}
\end{equation}%
which transforms like as $S\rightarrow \Omega S$. With this scalar multiplet
one can form a new tetrad in terms of the gauge invariant construction, $%
S^{-1}eS$. So, choosing appropriately the flat space components in the $S$
field one can turn the tetrad axial part to\ zero and establish symmetry
between Greek and Latin spacetime indices \cite{ish1}.

The issue still lies in the fact that the tetrads in (\ref{lllll}), along
with a neutral component, also include the $SU(N)$ hyperflavored components,
making it impossible to treat them as standard vielbein fields that satisfy
the invertibility conditions (\ref{or}). As one can see, a strict limitation
on the form of tetrads exists for these conditions to be upheld. Indeed, let
them generally have the\ covariant $SL(2N,C)$\ form 
\begin{equation}
e_{\mu }=e_{\mu }^{aK}\gamma _{a}\lambda ^{K},\text{ }e_{\mu
}^{aK}e_{b}^{\mu K^{\prime }}=\Delta _{b}^{aKK^{\prime }},\text{ }e_{\mu
}^{aK}e_{a}^{\nu K^{\prime }}=\Delta _{\mu }^{\nu KK^{\prime }}  \label{or1}
\end{equation}%
with some still unspecified constructions $\Delta _{b}^{aKK^{\prime }}$ and $%
\Delta _{\mu }^{\nu KK^{\prime }}$ which for a pure gravity case should
satisfy the standard arrangement 
\begin{equation}
\Delta _{b}^{a00}=\delta _{b}^{a}\text{ , \ }\Delta _{\mu }^{\nu 00}=\delta
_{\mu }^{\nu }  \label{or2}
\end{equation}%
Then multiplying the conditions (\ref{or1}) by the tetrad multiplets $%
e_{\sigma }^{bK^{\prime \prime }}$and $e_{a}^{\sigma K^{\prime \prime }}$,
respectively, one come after simple calculations 
\begin{equation}
\text{\ }e_{\mu }^{aK}=\Delta _{b}^{aK0}e_{\mu }^{b0}\text{, \ }e_{a}^{\mu
K^{\prime }}=\Delta _{a}^{bK^{\prime }0}e_{b}^{\mu 0}  \label{or3}
\end{equation}%
that finally gives 
\begin{equation}
\Delta _{b}^{aKK^{\prime }}=\Delta _{c}^{aK0}\Delta _{b}^{cK^{\prime }0}
\label{or4}
\end{equation}%
and correspondingly 
\begin{equation}
\Delta _{\mu }^{\nu KK^{\prime }}=\Delta _{\sigma }^{\nu K0}\Delta _{\mu
}^{\sigma K^{\prime }0}  \label{or4'}
\end{equation}%
As a result, for the constant and multiplicative form of these functions one
unavoidably comes to the only possible solution%
\begin{equation}
e_{\mu }^{aK}e_{b}^{\mu K^{\prime }}=\Delta _{b}^{aKK^{\prime }}=\delta
_{b}^{a}\delta ^{K0}\delta ^{K^{\prime }0}\text{ , \ }e_{\mu
}^{aK}e_{a}^{\nu K^{\prime }}=\Delta _{\mu }^{\nu KK^{\prime }}=\delta _{\mu
}^{\nu }\delta ^{K0}\delta ^{K^{\prime }0}  \label{or5}
\end{equation}%
which essentially mirrors the pure gravity case. Consequently, the
invertibility condition for tetrads is consistent with the $SL(2N,C)$
symmetry only if they belong to its $SL(2,C)$ part rather than the entire
group 
\begin{equation}
e_{\mu }^{aK}=e_{\mu }^{a}\delta ^{K0}  \label{or6}
\end{equation}%
This may appear as result of the spontaneous violation $SL(2N,C)$ in the
tetrad sector, as we will argue later.

Once the $SL(2N,C)$ transformation (\ref{rt}) becomes local one also need,
as ever, to introduce the gauge field multiplet $I_{\mu }$\ transforming as
usual%
\begin{equation}
I_{\mu }\rightarrow \Omega I_{\mu }\Omega ^{-1}-\frac{1}{ig}(\partial _{\mu
}\Omega )\Omega ^{-1}  \label{gg}
\end{equation}%
thus providing the fermion multiplet by covariant derivative 
\begin{equation}
\partial _{\mu }\Psi \rightarrow D_{\mu }\Psi =\partial _{\mu }\Psi
+igI_{\mu }\Psi \text{ }  \label{cov}
\end{equation}%
with universal gauge coupling constant $g$ of the proposed hyperunification.
The $I_{\mu }$ multiplet includes in general the vector and axial-vector
field submultiplets, and also the tensor field submultiplet 
\begin{equation}
I_{\mu }=V_{\mu }+A_{\mu }+T_{\mu }=\frac{1}{2}\left( V_{\mu }^{k}+iA_{\mu
}^{k}\gamma _{5}\right) \lambda ^{k}+\frac{1}{4}T_{\mu \lbrack
ab]}^{K}\gamma ^{ab}\lambda ^{K}\text{ \ \ }(K=0,k)  \label{ggggg}
\end{equation}%
as follows from its decomposition to the flat spacetime component fields.
Just the tensor fields provide gravitational interaction in the $SL(2N,C)$
HUTs that, aside from the standard linear curvature Lagrangian for gravity (%
\ref{127}), includes the conventional quadratic strength terms for all gauge
field submultiplets involved. This, as mentioned, poses the crucial problem
how one can selectively suppress tensor field interaction in these terms if
the tensor fields are members of the same gauge multiplet $I_{\mu }$\ as
vector and axial-vector fields and, therefore, should interact with the same
coupling constant $g$. Fortunately, the filtering mechanism described above
for the pure gravity case allows for a natural combination of the strong
internal symmetry forces related to the vector fields with the tiny
quadratic curvature gravity.

\subsection{Filtering with standard tetrads}

We begin by considering the tetrad filtering condition applied directly to
the general gauge multiplet itself (\ref{ggggg}) that could make the nature
of this condition clearer. Essentially, we impose the covariant constraint
of the form%
\begin{equation}
I_{\mu }=e_{\sigma }I_{\mu }e^{\sigma }/4  \label{c1}
\end{equation}%
using the "neutral" tetrads (\ref{or6}) which are only permitted in the
theory\footnote{%
This constraint when properly reiterated converts to an infinite series of
constraints%
\begin{equation*}
I_{\mu }=e_{\sigma _{1}}...e_{\sigma _{s}}I_{\mu }e^{\sigma
_{s}}...e^{\sigma _{1}}/4^{s}\text{ }(s=1,2,...)
\end{equation*}%
which all lead to the same outcome though.}. This yields%
\begin{equation}
I_{\mu }=\frac{1}{4}e_{\sigma }^{aK}e_{b}^{\sigma K^{\prime }}(\gamma
_{a}\lambda ^{K}I_{\mu }\gamma ^{b}\lambda ^{K^{\prime }})  \label{c2}
\end{equation}%
which, upon employing the invertibility conditions of the tetrad (\ref{or5}%
), results in the equality 
\begin{equation}
\frac{1}{2}\left( V_{\mu }^{k}+iA_{\mu }^{k}\gamma _{5}\right) \lambda ^{k}+%
\frac{1}{4}T_{\mu \lbrack ab]}^{K}\gamma ^{ab}\lambda ^{K}=\frac{1}{2}\left(
V_{\mu }^{k}-iA_{\mu }^{k}\gamma _{5}\right) \lambda ^{k}  \label{c2'}
\end{equation}%
This implies that the reduced gauge multiplet (\ref{c2}) comprises solely
the vector fields%
\begin{equation}
I_{\mu }=V_{\mu }^{k}\lambda ^{k}/2  \label{c3}
\end{equation}%
while the axial-vector and tensor field submultiplets vanish identically.
Remarkably, by imposing the covariant constraint (\ref{c1}) the starting $%
SL(2N,C)$ symmetry group is effectively reduced to the pure unitary $SU(N)$
symmetry case. In a sense, the constraint acts as a symmetry-breaking
mechanism, but unlike typical scenarios, nothing remains of the original $%
SL(2N,C)$ gauge sector except its $SU(N)$ part.

Now, as claimed, we propose that the gauge field multiplet $I_{\mu }$ (\ref%
{ggggg}) is "originated" from the prototype nondynamical multiplet $\mathcal{%
I}_{\mu }$ as per the underlying filtering condition (\ref{is}). Their
connection in the $SL(2N,C)$ symmetry framework acquires the form 
\begin{equation}
I_{\mu }=\frac{1}{4}e_{\sigma }\mathcal{I}_{\mu }e^{\sigma }=\frac{1}{4}%
e_{\sigma }^{aK}e_{b}^{\sigma K^{\prime }}(\gamma _{a}\lambda ^{K}\mathcal{I}%
_{\mu }\gamma ^{b}\lambda ^{K^{\prime }})  \label{im}
\end{equation}%
utilizing the "neutral" tetrads (\ref{or6}) permitted in the theory.

The point is, however, that again, as in the pure gravity case (\ref{eem}),
the tensor field submultiplet $T_{\mu \lbrack ab]}^{K}$ completely disappear
when filtered by tetrads satisfying the standard invertibility conditions (%
\ref{or5}). Indeed, using them one immediately comes to the filtering
relation%
\begin{equation}
I_{\mu }=\frac{1}{4}e_{\sigma }\mathcal{I}_{\mu }e^{\sigma }=\left( \mathcal{%
V}_{\mu }^{k}-i\mathcal{A}_{\mu }^{k}\gamma _{5}\right) \lambda ^{k}/2
\label{iki}
\end{equation}%
showing that, while tensor field submultiplet is cancelled, the vector and
axial-vector ones practically remain in the gauge multiplet (\ref{ggggg})
except that the axial-vector fields change the sign. In principle, one could
use it to eliminate them from the theory as well. This simply follows when
one adds to the condition (\ref{im}) the double filtering term as well 
\begin{equation}
I_{\mu }=\frac{1}{8}\left( e_{\sigma }\mathcal{I}_{\mu }e^{\sigma }+e_{\rho
}e_{\sigma }\mathcal{I}_{\mu }e^{\sigma }e^{\rho }/4\right) =\mathcal{V}%
_{\mu }^{l}\lambda ^{l}/2  \label{i6}
\end{equation}%
Comparing this with (\ref{c3}) we can conclude that such filtering is
equivalent to the case when the constraint is applied directly to the gauge
multiplet itself (\ref{c1}). However, in contrast, the prototype multiplet $%
\mathcal{I}_{\mu }$ appears to be free from any constraint. Remarkably, on
one hand, for the filtering case (\ref{i6}), one can covariantly transition
from the original $SL(2N,C)$ group to the $SU(N)$ symmetry gauged by the
vector fields. However, on the other hand, the tensor fields that could
induce gravity also disappear from the theory.

\subsection{Filtering with modified tetrads}

For incorporation of tensor fields in the gauge sector, it is necessary, as
in the pure gravity scenario discussed in Section 2.2, to go to tetrads
which are not strictly invertible. In such a scenario, the tensor field
multiplet will emerge within the filtered gauge multiplet $I_{\mu }$ once
the invertibility conditions turns out to be slightly shifted 
\begin{equation}
e_{\mu }^{aK}e_{b}^{\mu K^{\prime }}=(\delta _{b}^{a}+\varepsilon
q_{b}^{a})\delta ^{K0}\delta ^{K^{\prime }0}\text{ , \ }e_{\mu
}^{aK}e_{a}^{\nu K^{\prime }}=(\delta _{\mu }^{\nu }+\varepsilon p_{\mu
}^{\nu })\delta ^{K0}\delta ^{K^{\prime }0}  \label{or7}
\end{equation}%
being determined again by the tiny tensors $\varepsilon q_{b}^{a}$ and $%
\varepsilon p_{\mu }^{\nu }$ ($\varepsilon \ll 1$), respectively.
Analogously, the metric tensor deviation in the general $SL(2N,C)$ follows
from a similar equation%
\begin{equation}
e_{\mu }^{aK}e_{a\nu }^{K^{\prime }}=(g_{\mu \nu }+r_{\mu \nu })\delta
^{K0}\delta ^{K^{\prime }0}  \label{e7}
\end{equation}%
that actually works for neutral components%
\begin{equation}
e_{\mu }^{a0}e_{a\nu }^{0}=g_{\mu \nu }+r_{\mu \nu }  \label{e5}
\end{equation}%
and finally gives for the metric tensor%
\begin{equation}
g_{\mu \nu }\simeq e_{\mu }^{a0}e_{b\nu }^{0}[\delta _{a}^{b}-\varepsilon
q_{a}^{b}+\varepsilon ^{2}(qq)_{a}^{b}]\text{ }  \label{e6}
\end{equation}%
being identical to that in the pure gravity case (\ref{rr22}).

Actually, tetrads can again be regarded as those which are condensed, thus
having the form 
\begin{equation}
e_{\mu }^{aK}=(\delta _{\mu }^{a}h+\widehat{e}_{\mu }^{a0})\delta ^{K0},%
\text{ }\delta _{a}^{\mu }\widehat{e}_{\mu }^{a0}=0  \label{mai2}
\end{equation}%
where the effective Higgs mode\ $h$\ and zero-modes $\widehat{e}_{\mu }^{a0}$
are similar those given above in conditions (\ref{bd}-\ref{vev}). These
conditions suggest now a spontaneous-like breakdown of the local $SL(2N,C)$
symmetry for tetrads even though they are not considered as dynamical fields
in the theory.

The form (\ref{mai2}) remains, in fact, the original pure gravity case form
for neutral tetrad component, while the $SU(N)$ flavored tetrad components
are absent in the theory.

Note that a right choice of tetrad components is of primary importance
since, as in the above pure gravity case, just the tetrad-filtered gauge
multiplet is proposed to operate in the extended $SL(2N,C)$ theory. As a
consequence of the modified invertibility conditions for the tetrads taken
above (\ref{or7}), one has, after using the identity 
\begin{equation}
q_{b}^{a}\gamma _{a}\gamma ^{b}=q^{ab}\gamma _{a}\gamma _{b}=q^{ab}\eta
_{ab}=0  \label{qs}
\end{equation}%
for the symmetrical and traceless tensor $q^{ab}$ (\ref{q'}), a unique form
for the filtered gauge multiplet $I_{\mu }$\ 
\begin{subequations}
\begin{equation}
I_{\mu }=\left( \mathcal{V}_{\mu }^{k}-i\mathcal{A}_{\mu }^{k}\gamma
_{5}\right) \lambda ^{k}/2+\frac{\varepsilon }{4}\mathbf{T}_{\mu \lbrack
ab]}^{K}\gamma ^{ab}\lambda ^{K}\text{ }  \label{BM}
\end{equation}%
It is noteworthy that, in contrast to the standard tetrad case (\ref{iki}),
it also contains the tensor field submultiplet which is given by an
expression 
\end{subequations}
\begin{subequations}
\begin{equation}
\mathbf{T}_{\mu \lbrack ab]}^{K}=(\mathcal{T}_{\mu \lbrack bc]}^{K}q_{a}^{c}-%
\mathcal{T}_{\mu \lbrack ac]}^{K}q_{b}^{c})/2  \label{BM1}
\end{equation}%
This actually extends the pure gravity case (\ref{bb}) in a sense that the
hyperflavored tensor field components\ $\mathbf{T}_{\mu \lbrack ab]}^{k}$\
also come into play. However, as we see later they appear insignificant only
contributing into the tiny four-fermion (spin current-current) interaction
in the fermion matter sector.

As one can readily confirm, the vector and axial-vector submultiplets in the
gauge multiplet (\ref{ggggg}) remain unaffected during the filtering process
(unless the special constraint (\ref{c1}) excluding the axial-vector field
in the theory is applied). Meanwhile, the tensor field components appear
again, as in the pure gravity case, to be completely controlled by the
tetrad invertibility deviations $\varepsilon q_{b}^{a}$ and, therefore, are
significantly weakened. The product of the deviation tensor $q_{b}^{a}$ with
the prototype tensor fields $\mathcal{T}_{\mu \lbrack ab]}^{K}$ defines the
new tensor field multiplet $\mathbf{T}_{\mu \lbrack ab]}^{K}$ (\ref{BM1}),
through which and tiny parameter $\varepsilon $ the theory is ultimately
expressed. The hyperunification of the basic elementary forces in this
theory does not preclude the tensor field submultiplet from having the
vanishingly small quadratic strength terms being scaled as $\varepsilon ^{2}$
and, as a consequence, can be neglected. Therefore, as we confirm below, the
final theory tends to the conventional Einstein-Cartan type theory for
gravity coupled with the gauge $SU(N)$ theories for other interactions.

\subsection{Gravity inducing tensor fields}

Let us now construct the field strength for the gauge multiplet $I_{\mu }$
in the $SL(2N,C)$ hyperunified theory 
\end{subequations}
\begin{equation}
I_{\mu \nu }=\partial _{\lbrack \mu }I_{\nu ]}+ig[I_{\mu },I_{\nu
}]=(V+A)_{\mu \nu }+T_{\mu \nu }  \label{i}
\end{equation}%
which includes the terms corresponding the vector, axial-vector and tensor
field submultiplets, respectively. Expressed through the prototype $\mathcal{%
I}_{\mu }$ multiplet components according to the taken filtered form (\ref%
{BM}), this strength tensor comes to 
\begin{eqnarray}
I_{\mu \nu } &=&\frac{1}{2}\partial _{\lbrack \mu }\left( \mathcal{V}^{k}-i%
\mathcal{A}^{k}\gamma _{5}\right) _{\nu ]}\lambda ^{k}-\frac{1}{2}%
f^{ijk}g\left( \mathcal{V}^{i}-i\mathcal{A}^{i}\gamma _{5}\right) _{\mu }(%
\mathcal{V}^{j}-i\mathcal{A}^{j}\gamma _{5})_{\nu }\lambda ^{k}  \notag \\
&&+\frac{\varepsilon }{4}\left( \partial _{\lbrack \mu }\mathbf{T}_{\nu
]}^{[ab]K}\gamma _{ab}\lambda ^{K}+i\frac{\varepsilon g}{4}\mathbf{T}_{\mu
}^{[ab]K}\mathbf{T}_{\nu }^{[a^{\prime }b^{\prime }]K^{\prime }}[\lambda
^{K}\gamma _{ab},\lambda ^{K^{\prime }}\gamma _{a^{\prime }b^{\prime
}}]\right)  \label{f}
\end{eqnarray}%
Similarly, the gauge invariant fermion matter couplings, when given in terms
of the $\mathcal{I}_{\mu }$\ submultiplets, take the form%
\begin{equation}
e\mathcal{L}_{M}=-\frac{g}{2}\overline{\Psi }\left\{ e^{\mu },\left[ \frac{1%
}{2}\left( \mathcal{V}_{\mu }^{k}-i\mathcal{A}_{\mu }^{k}\gamma _{5}\right)
\lambda ^{k}+\frac{\varepsilon }{16}\mathbf{T}_{\mu \lbrack ab]}^{K}\gamma
^{ab}\lambda ^{K}\right] \right\} \Psi  \label{lfm}
\end{equation}

As one can readily observe, the vector and axial-vector fields interact
everywhere in (\ref{f}) and (\ref{lfm}) with the universal gauge coupling
constant $g$ of $SL(2N,C)$. In contrast, all tensor field terms incorporate
the aforementioned tiny parameter $\varepsilon $. Consequently, the total\
Lagrangian will contain the conventional quadratic strength terms for the
vector and axial-vector fields, while in the first order in the parameter $%
\varepsilon $ only the linear strength terms of the tensor fields emerge,
alongside the fermion matter couplings. The quadratic strength terms of the
tensor field multiplet, which would provide its propagation and an ordinary
gauge interaction, scale as $\varepsilon ^{2}$ and are therefore negligible
in this regime.

Leaving aside for the moment vector and axial-vector fields, we now focus on
the hyperunified gravity Lagrangian taken in the Palatini type form%
\begin{equation}
e\mathcal{L}_{G}\sim Tr\{[e^{\mu },e^{\nu }]I_{\mu \nu }\}  \label{127a}
\end{equation}%
In the $SL(2N,C)$ case the strength tensor $I_{\mu \nu }$, apart from tensor
submultiplet, comprises the vector and axial-vector submultiplets as well.
However, due to the neutral tetrad chosen (\ref{or6}) and their commutator
given by%
\begin{equation}
\lbrack e^{\mu },e^{\nu }]=-2ie_{a}^{\mu 0}e_{b}^{\nu 0}\gamma ^{ab}
\label{Tr2}
\end{equation}%
one can easily confirm that the they do not contribute to the gravity
Lagrangian (\ref{127a}).

Eventually, for the tensor field strength in (\ref{f}) one has, after taking
the necessary traces of products involving $\gamma $ and $\lambda $
matrices, the following gravity Lagrangian 
\begin{equation}
e\mathcal{L}_{G}=\frac{1}{2\kappa }\left( \partial _{\lbrack \mu }\mathbf{T}%
_{\nu ]}^{[ab]0}+\varepsilon g\eta _{cd}\mathbf{T}_{[\mu }^{[ac]K}\mathbf{T}%
_{\nu ]}^{[bd]K}\right) e_{a}^{\mu 0}e_{b}^{\nu 0}  \label{p2}
\end{equation}%
The gravity constant $\kappa $ is assumed to absorb here one power of the
tiny parameter $\varepsilon $ (along with the factor $N$ associated with the
internal $U(N)$ symmetry). Consequently, the new tensor fields $\mathbf{T}%
_{\mu }^{[ab]K}$ appears in the Lagrangian $e\mathcal{L}_{G}$ with the
effective coupling constant $\varepsilon g$ rather than $g$, as the vector
and axial-vector fields do in their own quadratic strength Lagrangians. In a
similar way, one has for the fermion matter Lagrangian of the tensor
submultiplet in (\ref{lfm}) 
\begin{equation}
e\mathcal{L}_{M}^{(T)}=-\frac{\varepsilon g}{2}\epsilon ^{abcd}\mathbf{T}%
_{\mu \lbrack ab]}^{K}\overline{\Psi }e_{c}^{\mu 0}\gamma _{d}\lambda
^{K}\gamma ^{5}\Psi  \label{m1}
\end{equation}%
where couplings of the tensor fields with the neutral and flavored spin
density currents also appear with the same effective coupling constant $%
\varepsilon g$.

Notably, in the Lagrangian (\ref{p2}), there are only kinetic terms for the
neutral tensor field component $\mathbf{T}_{\mu }^{[ab]0}$, while the
interaction terms contain the entire $U(N)$ multiplet $\mathbf{T}_{[\mu
}^{[ab]K}$. This implies that only the neutral tensor field truly gauges
gravity, while the $SU(N)$\ flavored ones $\mathbf{T}_{\mu }^{[ab]k}$ are
simply given by the corresponding spin currents $\epsilon ^{abcd}\overline{%
\Psi }e_{\mu c}^{0}\gamma _{d}\lambda ^{k}\gamma ^{5}\Psi $. When they both, 
$\mathbf{T}_{[\mu }^{[ab]0}$ and $\mathbf{T}_{\mu }^{[ab]k}$, are
independently eliminated from the entire tensor field Lagrangian $e\mathcal{L%
}_{G}+e\mathcal{L}_{M}^{(T)}$, one arrives at the Einstein-Cartan type
gravity containing, besides the usual GR, the tiny 4-fermion spin density
interaction 
\begin{equation}
\varepsilon \kappa \left( \overline{\Psi }\gamma _{c}\gamma ^{5}\lambda
^{K}\Psi \right) (\overline{\Psi }\gamma ^{c}\gamma ^{5}\lambda ^{K}\Psi )
\label{6}
\end{equation}%
which in contrast to the standard case \cite{Kibble} includes the flavored
four-fermion interaction terms as well, albeit further weakened by the small
parameter $\varepsilon $.

We have already mentioned in Section 2.2 that the weakness of the tensor
field induced gravity may look fictitious since it might be circumvented by
a simple rescaling of the field itself. However, as follows, this rescaling
trick only works for pure gauge gravity with local $SL(2,C)$ symmetry. In
the framework of $SL(2N,C)$ gauge theory, where the tensor field is proposed
to be unified with ordinary vector and axial-vector fields which interact
with the $O(1)$ coupling constants, rescaling is not an option. \ The point
is that the weakness of the filtered tensor field is exclusive to the tensor
field submultiplet and does not extend to the entire gauge multiplet $I_{\mu
}$ (\ref{BM}) of $SL(2N,C)$. As a result, it cannot be overcome by mere
rescaling. Indeed, rescaling of only the tensor field submultiplet $\mathbf{T%
}_{\mu }=\mathbf{T}_{\mu }^{\prime }/\varepsilon $\ inside\ $I_{\mu }$ is
not permitted by the $SL(2N,C)$ gauge invariance in the theory, nor by the
filtering condition (\ref{is}) itself. Conversely, rescaling of the entire
multiplet $I_{\mu }=I_{\mu }^{\prime }/\varepsilon $, while "normalizing"
the tensor field part in total Lagrangian, will "denormalize" the vector and
axial-vector field terms,

\begin{equation}
\mathcal{L}(\mathcal{V},\mathcal{A,}\varepsilon \mathbf{T)}=\mathcal{L}%
^{\prime }(\mathcal{V}^{\prime }/\varepsilon ,\mathcal{A}^{\prime
}/\varepsilon ,\mathbf{T}^{\prime })  \label{ll'}
\end{equation}%
This necessitates a new rescaling of the entire multiplet $I_{\mu }^{\prime
}=\varepsilon I_{\mu }^{\prime \prime }$ which bring us back to the initial
point with the tiny filtered tensor field. Thus, both cases render the
rescaling scenarios untenable.

\subsection{Hyperflavor mediating vector fields}

Turn now to the vector and axial-vector fields which are the basic spin-1
carriers of the hyperflavor $SU(N)$ symmetry in the $SL(2N,C)$ theory. Their
own sector stemming from the common strength tensor (\ref{f}) looks as%
\begin{equation}
e\mathcal{L}^{(VA)}=-\frac{1}{4}[\partial _{\lbrack \mu }\mathcal{V}_{\nu
]}^{k}-gf^{ijk}(\mathcal{V}_{\mu }^{i}\mathcal{V}_{\nu }^{j}+\mathcal{A}%
_{\mu }^{i}\mathcal{A}_{\nu }^{j})]^{2}-\frac{1}{4}[\partial _{\lbrack \mu }%
\mathcal{A}_{\nu ]}^{k}]^{2}  \label{I}
\end{equation}%
where, as one can see, the vector fields acquire a conventional gauge theory
form, while the axial-vector field couplings break this gauge invariance. At
the same time, as follows from the matter sector of the theory (\ref{lfm}),
the vector fields interact with ordinary matter fermions%
\begin{equation}
e\mathcal{L}_{M}^{(V)}=-\frac{g}{2}\mathcal{V}_{\mu }^{i}\overline{\Psi }%
e_{a}^{\mu 0}\gamma ^{a}\lambda ^{i}\Psi  \label{m3}
\end{equation}%
while axial-vector fields do not, thus being sterile to them.

Generally, one could try to adapt the axial-vector fields to reality though
there is no sign of they actually existing. The traditional way would be to
make these axial-vector fields superheavy through some enormously extended
Higgs sector remaining, at the same time, vector fields gauging the Standard
Model massless or enough light. This seems to be quite difficult since the
axial-vector fields want to follow the same pattern of the mass formation as
the vector fields do. Anyway, despite the gauge $SL(2N,C)$ invariance in the
theory, the very presence of the axial-vector fields breaks the gauge $SU(N)$
invariance related to the vector fields, thus only leaving the global $SU(N)$
symmetry in the theory.

In this connection, a rather interesting way could be if these axial-vector
fields were condensed at some Planck order scale $\mathcal{M}$, thus
providing a true vacuum in the theory, $\left\langle \mathcal{A}_{\mu
}^{i}\right\rangle =\mathfrak{n}_{\mu }^{i}\mathcal{M}$, whose direction is
given by the unit Lorentz vector $\mathfrak{n}_{\mu }^{i}$. Remarkably, in
this vacuum, as is directly seen, gauge invariance for the vector fields is
completely restored, though a tiny spontaneous breaking of the Lorentz
invariance at the scale $\mathcal{M}$ may appear. More details about such a
possibility can be found in the recent paper \cite{jpl}.

A more radical approach, as mentioned earlier, would be to impose the
specific filtering condition (\ref{i6}) which essentially corresponds to an
existence the constraint in the theory being applied directly to the gauge
multiplet itself (\ref{c1}). This filtering condition automatically excludes
the axial-vector and tensor fields from the gauge multiplet\ $I_{\mu }$ when
one uses the standard tetrads. However, for the modified tetrads, that
satisfy the slightly shifted invertibility conditions (\ref{or7}), one
ultimately obtains, in contrast to (\ref{i6}), the total gauge multiplet $%
I_{\mu }$ having the form 
\begin{equation}
I_{\mu }=\mathcal{V}_{\mu }^{l}\lambda ^{l}/2+\frac{\varepsilon }{8}\mathbf{T%
}_{\mu \lbrack ab]}^{K}\gamma ^{ab}\lambda ^{K}+O(\varepsilon ^{2})
\label{i10}
\end{equation}%
This multiplet, in addition to the flavor-mediating vector gauge fields of
the local $SU(N)$ symmetry, incorporates the appropriately damped tensor
fields that underlie the gravity sector in the hyperunified theory (terms of
order $\varepsilon 
%TCIMACRO{\U{b2}}%
%BeginExpansion
{{}^2}%
%EndExpansion
$ are disregarded for simplicity).

\subsection{Symmetry breaking scenario}

It is clear that hyperunification of all elementary forces supposes that,
while gravity may have an unique linear tensor field strength Lagrangian $%
L_{G}$ of the type given in (\ref{p2}), the quadratic strength terms of
other components of gauge multiplet $I_{\mu }$ are naturally unified in
their common $SL(2N,C)$ invariant Lagrangian. The tiny quadratic terms for
the tensor fields containing the $O(\varepsilon ^{2})$ and higher order
terms appear unessential compared to the strength bilinears for vector
fields.\ So, hyperunification definitely rules out the quadratic curvature
terms for tensor fields in the filtered $SL(2N,C)$ gauge theory. The only
place where tensor field submultiplet manifests itself is the $SL(2,C)$
gravity gauged by its tiny neutral component $\varepsilon \mathbf{T}_{\mu
\lbrack ab]}^{0}$. Accordingly, one eventually has for a gauge sector of the
unified Lagrangian 
\begin{equation}
eL_{U}=eL_{G}-\frac{1}{4}V_{\mu \nu }^{k}V^{\mu \nu k}  \label{r67''}
\end{equation}%
containing, apart from the Einstein-Cartan type theory, the standard $SU(N)$
invariant vector field part (provided that the axial-vector fields are
properly filtered out of the theory in the way described above).\ \ \ \ \ \
\ \ \ \ \ \ \ \ \ \ 

A conventional breaking scenario of the $SL(2N,C)$ invariance in the theory
would depend in general on the proper set of scalar fields which could break
this invariance first to the intermediate $SU(N)\times SL(2,C)$ symmetry and
then to the Standard Model. In our case, however, one does not need to cause
the first stage of symmetry breaking since, as is readily seen in (\ref{i10}%
), all the gauge submultiplets related to the "nondiagonal" generators of $%
SL(2N,C)$ are properly weakened (tensor fields) or completely filtered out
of the theory (axial-vector fields).

As to the internal $SU(N)$ symmetry violation down to the Standard Model one
actually need to have the adjoint scalar multiplets of the type 
\begin{equation}
\Phi =(\phi ^{K}+i\phi _{5}^{K}\gamma _{5}+\phi _{ab}^{K}\gamma
^{ab}/2)\lambda ^{K}  \label{ad}
\end{equation}%
which transform under $SL(2N,C)$ as%
\begin{equation}
\Phi \rightarrow \Omega \Phi \Omega ^{-1}\text{\ }  \label{fo}
\end{equation}%
It generally contains, apart the scalar components, the pseudoscalar and
tensor components as well. However, as in the above gauge multiplet case,
one can use again the tetrad projection mechanism to filter away these
"superfluous" components, just like as it was done in (\ref{i6})%
\begin{equation}
\Phi =\frac{1}{8}(e_{\sigma }\boldsymbol{\Phi }e^{\sigma }+e_{\rho
}e_{\sigma }\mathbf{\Phi }e^{\sigma }e^{\rho }/4)  \notag
\end{equation}%
where $\mathbf{\Phi }$ is some prototype scalar field multiplet. As a
result, with tetrads satisfying the invertibility conditions (\ref{or7})
there is only left the pure scalar components in the $SU(N)$ symmetry
breaking multiplet $\Phi $ 
\begin{equation}
\Phi =\mathbf{\phi }^{k}\lambda ^{k}\text{ \ \ \ \ }(k=1,...,N^{2}-1)\text{\
\ }  \notag
\end{equation}%
providing (with other similar scalar multiplets) the breaking of the $SU(N)$
GUT down to the Standard Model. The final symmetry breaking to $%
SU(3)_{c}\times U(1)_{em}$ is provided by extra scalar multiplets whose
assignment depends on which multiplets are chosen for quarks and leptons.

\subsection{Final remarks}

We have observed that the filtered tensor field only manifests when the
invertibility conditions for tetrads are appropriately shifted, as argued
earlier in (\ref{or7}). This, in turn, leads to a slight modification of the
metric tensor (\ref{e6}), indicating a tiny departure from general
covariance, albeit in a controllable manner determined by the minute
parameter $\varepsilon $.

Nevertheless, it is conceivable to manage the filtering mechanism in a way
that preserves general covariance. This could only be achieved if there
exist two types of tetrads satisfying the invertibility conditions 
\begin{equation}
e_{\mu }^{aK}e_{b}^{\mu K^{\prime }}=\delta _{b}^{a}\delta ^{K0}\delta
^{K^{\prime }0},\text{ \ }E_{\mu }^{aK}E_{b}^{\mu K^{\prime }}=(\delta
_{b}^{a}+\varepsilon q_{b}^{a})\delta ^{K0}\delta ^{K^{\prime }0}  \label{eE}
\end{equation}%
where the primary tetrad $e$ adheres to the standard invertibility
condition, while the auxiliary tetrad $E$ exhibits a slight
non-invertibility deviation $\varepsilon q_{b}^{a}$. One can additionally
require the entire Lagrangian to remain invariant under the reflection
transformation 
\begin{equation}
e\rightarrow e\text{ , \ }E\rightarrow -E\text{\ \ }  \label{ee1}
\end{equation}%
This ensures that the tetrad $E$, in contrast to the primary tetrad $e$
solely parametrizes the spacetime background and does not participate in the
matter fermion terms. Consequently, while the tetrad $E$ essentially
determines the filtering condition $I_{\mu }=E_{\sigma }\mathcal{I}_{\mu
}E^{\sigma }/4$, a vanishingly small violation of general covariance, caused
by its approximate invertibility condition in (\ref{eE}), appears beyond the
scope of the Einstein-Cartan gravity emerged.

In this context, there are indeed numerous instances in physics where the
consideration of a manifold equipped with two distinct vielbein fields
becomes necessary. This notably occurs in bimetric theories, where two
disparate metrics are defined on the same spacetime manifold, including the
case when one of the metrics is nondynamical \cite{ish2}. Importantly, this
concept extends to certain formulations of bigravity theory, which could
potentially serve as a base for the massive gravity \cite{gab, has}.

Nevertheless, despite the possibility of employing two metrics within the
framework of the $SL(2N,C)$ gauge theory, the case with the single metric,
even if general covariance is slightly broken, appears to be more
economically viable. Consequently, we have proceeded with this version in
our hyperunified model.

\section{From\ hyperunification to\ GUTs}

\subsection{$SU(5)$ and its direct\ extensions}

Let us now consider more closely how the $SL(2N,C)$ type model can be
applied to some known GUTs starting with a conventional $SU(5)$ \cite{gg}
which would stem from the $SL(10,C)$ HUT. In this case some of its
low-dimensional multiplets of the chiral (lefthanded for certainty) fermions
can be given in terms of the $SU(5)\times SL(2,C)$ components as 
\begin{eqnarray}
&&\Psi _{L}^{i\boldsymbol{a}}\text{ },\text{\ \ \ \ \ }10=(\overline{5},2)%
\text{ }  \label{sl1} \\
\Psi _{L[\boldsymbol{a}i,\text{ }j\boldsymbol{b}]} &=&\Psi _{L[ij]\{%
\boldsymbol{ab}\}}+\Psi _{L\{ij\}[\boldsymbol{ab}]},\text{ \ }45=(10,\text{ }%
3)+(15,\text{ }1)\text{ }  \label{sl2}
\end{eqnarray}%
where we have used that a common antisymmetry on two or more joint $SL(10,C)$
indices ($i\boldsymbol{a}$,\ $j\boldsymbol{b}$, $k\boldsymbol{c}$) means
antisymmetry in the internal indices ($i,j,k=1,...,5$) and symmetry in the
chiral spinor ones ($\boldsymbol{a},\boldsymbol{b},\boldsymbol{c}=1,2$), and
vice versa (dimension of representations are also indicated). One can see
that, while the $SU(5)$ antiquintet can easily be constructed (\ref{sl1}),
its decuplet is not contained in the pure antisymmetric $SL(10,C)$
representations (\ref{sl2}). Moreover, the tensor (\ref{sl2}) corresponds in
fact to the collection of vector and scalar multiplets rather than the
fermion ones.

Note in this connection, that all GUTs where fermions are assigned to the
pure antisymmetric representations seem to be also irrelevant since the spin
magnitude of appearing states are not in conformity with what we have in
reality. The most known example of this kind is the $SU(11)$ GUT \cite{ge}
with all three quark-lepton families collected in its one-, two-, three-,
and four-index antisymmetric representations. No doubt this GUT should also
be excluded in the framework of the considered $SL(2N,C)$ theories.
Actually, for the right $1/2$ spin value of ordinary quarks and leptons
these theories should include more complicated fermion multiplets having in
general the upper and lower indices rather than the pure asymmetric ones.
The point is, however, that such multiplets appear enormously large and
contain in general lots of exotic states which never been detected. This
could motivate to seek a possible solution in the composite nature of quarks
and leptons for whose constituents -- preons -- the $SL(2N,C)$ unification
might look much simpler.

\subsection{$SU(8)$ with composite quarks and leptons}

Following the recent discussion \cite{ch}, we introduce $N$ lefthanded and $%
N $\ righthanded preons being the fundamental multiplets $P_{Li\boldsymbol{a}%
\text{ }}^{\alpha }$ and $P_{Ri\boldsymbol{a}\text{ }}^{\alpha ^{\prime }}$
of the vectorlike "metaflavor" $SL(2N,C)$ HUT\ symmetry ($i=1,...,N;$ $%
\boldsymbol{a}=1,2$) times some local left-right "metacolor" $%
SO(n)_{L}\times SO(n)_{R}$ symmetry ($\alpha =$ {$1,...,n$}; $\alpha
^{\prime }=1,...,n$) binding preons inside quarks and leptons\footnote{%
By tradition, we call them the "metaflavor" and "metacolor" symmetry, while
still referring to the $SU(N)$ subgroup of $SL(2N,C)$ as the hyperflavor
symmetry.}. Both of these symmetries are obviously anomaly-free and the
numbers of metaflavors ($N$) and metacolors ($n$) are not yet determined.
The metaflavor symmetry describes preons at small distances as well as their
composites at large ones. They are produced individually from the lefthanded
and righthanded preons due to confining forces of the above metacolor
symmetry. Some of these composites, including the observed quarks and
leptons, are expected to be much lighter than their composition scale. For
that, the accompanying chiral symmetry $SU(N)_{L}\times SU(N)_{R}$ of the
preons should be preserved at large distances in a way that -- when it is
considered as the would-be local symmetry group with some spectator gauge
fields and fermions -- the corresponding triangle anomaly matching
conditions \cite{t} are satisfied. Namely, the $SU(N)_{L}^{3}\ $ and $%
SU(N)_{R}^{3}$ anomalies related to $N$ lefthanded and $N$ righthanded
preons have to individually match those for lefthanded and righthanded
composite fermions being produced by the $SO(n)_{L}$ and $SO(n)_{R}$
metacolor forces, respectively.

Moreover, as is turned out, just this condition, when being properly
strengthened, can determine the particular metaflavor symmetry $SL(2N,C)$ in
the theory. Indeed, we first assume that all composites, both lefthanded and
righthanded, have just the three-preon configuration ($n=3$), thus fixing
the metacolor symmetry to $SO(3)_{L}\times SO(3)_{R}$. And second and most
importantly, they belong to a single representation of their chiral
symmetries $SU(N)_{L}$ and $SU(N)_{R}$, respectively, rather than to some
set of representations. Then it turns\ out that among all their third-rank
representations the anomaly matching condition\ holds individually\ only for
multiplets of the type $\psi _{\lbrack i\text{ }j]L}^{k}$ and $\psi
_{\lbrack i\text{ }j]R}^{k}$ ($i,j,k=1,2,...,N$), that gives the unique
solution to the\ number\ of\ preons $N$, both lefthanded\ and\ righthanded, 
\begin{equation}
N^{2}/2-7N/2-1=3,\text{ }N=8\text{ }  \label{8-1}
\end{equation}%
This means that among all possible chiral symmetries only the $%
SU(8)_{L}\times SU(8)_{R}$ symmetry can in principle provide masslessness of
lefthanded\ and\ righthanded fermion composites at large distances. This in
turn identifies -- among all metaflavor $SL(2N,C)$ symmetries -- just $%
SL(16,C)$ as the most likely candidate for hyperunification. Note that, in
contrast to the above global chiral symmetry, in the local $SL(16,C)$
metaflavor theory, being as yet vectorlike, all metaflavor triangle
anomalies are automatically cancelled out.

Turning now from the chiral symmetry multiplets $\psi _{\lbrack i\text{ }%
j]L,R}^{k}$ to the corresponding\ $SL(16,C)$ composite multiplets $\Psi
_{\lbrack i\boldsymbol{a},\text{ }j\boldsymbol{b}]L,R}^{k\boldsymbol{c}}$
one can write them in terms of the $SU(8)\times SL(2,C)$ components as the
collection 
\begin{equation}
\Psi _{\lbrack i\boldsymbol{a},\text{ }j\boldsymbol{b}]}^{k\boldsymbol{c}%
}=\Psi _{\lbrack ij]\{\boldsymbol{ab}\}}^{k\boldsymbol{c}}+\Psi _{\{ij\}[%
\boldsymbol{ab}]}^{k\boldsymbol{c}},\text{ \ }1904=(216,2)+(216+8,\text{ }%
4)+(280+8,\text{ }2)  \label{216C}
\end{equation}%
which contains some spin $1/2$ and $3/2$ lefthanded and righthanded
composite fermion submultiplets. Meanwhile, as one can easily confirm, among
all submultiplets in (\ref{216C}) only the $(216,2)_{L,R}$ ones satisfy
individually the anomaly matching condition for the chiral $SU(8)_{L}$ and $%
SU(8)_{R}$ symmetries, respectively. As a result, all the other
submultiplets there have then to acquire superheavy masses. This actually
means that only the $SU(8)\times SL(2,C)$ subgroup of the $SL(16,C)$ HUT
symmetry survives at large distances where the composite fermions emerge.
Surprisingly enough, this is consistent with what we had above, albeit from
a different perspective. Namely, the filtered $SL(2N,C)$ gauge theory, in
which only neutral tensor field and vector field multiplet remain, turns out
to be effectively reduced to the $SU(N)\times SL(2,C)$ invariant theory.
Now, in the composite model, for the particular case of metaflavored
symmetry $SL(16,C)$, this independently follows from the preservation of the
accompanying chiral symmetry $SU(8)_{L}\times SU(8)_{R}$ at large distances,
thus leading to the theory with the residual metaflavor symmetry $%
SU(8)\times SL(2,C)$.

Remarkably, the above $(216,2)_{L,R}$ submultiplets being decomposed into
the standard $SU(5)$ GUT and family symmetry $SU(3)_{F}$ looks as%
\begin{equation}
(216,2)_{L,R}=[(\overline{5}+10,\text{ }\overline{3}%
)+(45,1)+(5,8+1)+(24,3)+(1,3)+(1,\overline{6})]_{L,R}  \label{216}
\end{equation}%
where the first term in the squared brackets, when taken for lefthanded
states in $216_{L}$, describes all three quark-lepton families being the
family symmetry triplets. However, there are also the similar righthanded
states in $216_{R}$ in our still vectorlike $SL(16,C)$ theory. This means
that, while preons are left massless being protected by their own
metacolors, the composites (\ref{216}) being metacolor singlets could in
principle pair up and acquire the heavy Dirac masses.

To avoid this for the submultiplet of physical quarks and leptons in (\ref%
{216}), $(\overline{5}+10,$ $\overline{3})_{L}$, one may propose, following
the scenario developed in\ \cite{ch}, some spontaneous breaking of the basic 
$L$-$R$ symmetry in the theory. This is assumed to follow from the sector of
righthanded preons that reduces the chiral symmetry of their composites down
to $[SU(5)\times SU(3)]_{R}$. Actually, such a breaking may readily appear
due to a possible condensation of massive composite scalars which
unavoidably appear in the theory together with composite fermions. This
means that, though the massless righthanded preons still possess the $%
SU(8)_{R}$ symmetry, the masslessness of their composites at large distances
is now solely controlled by its remained $[SU(5)\times SU(3)]_{R}$ part.
Thus, while nothing really happens with the lefthanded preon composites
still completing the total multiplet $(216,2)_{L}$ in (\ref{216}), the
righthanded preon composites with their residual chiral symmetry no longer
include all submultiplets given in $(216,2)_{R}$. Very remarkably, the
corresponding anomaly matching condition "organizes" their composite
spectrum in such a way that the submultiplet $(\overline{5}+10,$ $\overline{3%
})_{R}$ is absent among the righthanded preon composites. As a result, all
the lefthanded submultiplets in $(216,2)_{L}$, except the $(\overline{5}+10,$
$\overline{3})_{L}$, will then pair up, thus becoming heavy and decoupling
from laboratory physics\ \cite{ch}.

Accordingly, once the $L$-$R$ symmetry is violated in the theory, the
vectorlike metaflavor symmetry $SU(8)\times SL(2,C)$, while still working
for preons, will also break down to its subgroup $[SU(5)\times
SU(3)_{F}]\times SL(2,C)$ for their large-distance composites. So, one
eventually comes to the conventional $SU(5)$ GUT \cite{gg} together with the
extra local $SU(3)_{F}$ family symmetry \cite{h} describing just three
standard families of composite quarks and leptons. Both types of the
triangle anomalies, $SU(5)^{3\text{ \ \ }}$and $SU(3)_{F}^{3}$, emerging at
this stage are properly cancelled out in the theory.

The further symmetry violation is related, as was mentioned above, to the
adjoint scalar field multiplet $\Phi $ (\ref{ad}) which in the present
context breaks the $SU(5)$ to the Standard Model. As to the final breaking
of the SM and accompanied family symmetry $SU(3)_{F}$, it appears through
the extra multiplets $H^{[i\boldsymbol{a},j\boldsymbol{b},k\boldsymbol{c},l%
\boldsymbol{d}]}$, and $\chi _{\lbrack i\boldsymbol{a},j\boldsymbol{b}]}$
and $\chi _{\{i\boldsymbol{a},j\boldsymbol{b}\}}$ of $SL(16,C)$,
respectively. These multiplets contain, among others, the true scalar
components which develop the corresponding VEVs and give masses to the weak
bosons, as well as the flavor bosons of the $SU(3)_{F}$. They also generate
masses to quarks and leptons located in the lefthanded fermion multiplet (%
\ref{216C}, \ref{216}) through the $SL(16,C)$ invariant Yukawa couplings%
\begin{eqnarray}
&&\frac{1}{\mathcal{M}}\left[ \Psi _{\lbrack j\boldsymbol{a},\text{ }k%
\boldsymbol{b}]L}^{i\boldsymbol{c}}C\Psi _{\lbrack m\boldsymbol{e},\text{ }n%
\boldsymbol{f}]L}^{l\boldsymbol{d}}\right] H^{\{[j\boldsymbol{a},k%
\boldsymbol{b]},[m\boldsymbol{e},n\boldsymbol{f}]\}}(a_{u}\chi _{\lbrack i%
\boldsymbol{c},l\boldsymbol{d}]}+b_{u}\chi _{\{i\boldsymbol{c},l\boldsymbol{d%
}\}})  \notag \\
&&\frac{1}{\mathcal{M}}\left[ \Psi _{\lbrack j\boldsymbol{a},\text{ }k%
\boldsymbol{b}]L}^{i\boldsymbol{c}}C\Psi _{\lbrack i\boldsymbol{c},\text{ }m%
\boldsymbol{e}]L}^{l\boldsymbol{d}}\right] H^{\{[j\boldsymbol{a},k%
\boldsymbol{b]},[m\boldsymbol{e},n\boldsymbol{f}]\}}(a_{d}\chi _{\lbrack l%
\boldsymbol{d},n\boldsymbol{f}]}+b_{d}\chi _{\{l\boldsymbol{d},n\boldsymbol{f%
}\}})  \label{yc}
\end{eqnarray}%
with different index contraction for the up quarks, and down quarks and
leptons, respectively ($i,j,k,l,m,n=1,...,8;$ $\boldsymbol{a},\boldsymbol{b},%
\boldsymbol{c},\boldsymbol{d},\boldsymbol{e},\boldsymbol{f}=1,2$). The mass $%
\mathcal{M}$ stands for some effective scale in the theory that in the
composite model of quarks and leptons can be related to their compositeness
scale, while $a_{u,d}$ and $b_{u,d}$ are some dimensionless constants of the
order of $1$. Actually, these couplings contain two types of scalar
multiplets with the following $SU(8)\times SL(2,C)$ components -- the $H$
multiplet $H^{\{[j\boldsymbol{a},k\boldsymbol{b]},[m\boldsymbol{e},n%
\boldsymbol{f}]\}}$ containing the true scalar components%
\begin{equation}
H^{[jkmn]\{[\boldsymbol{ab}],[\boldsymbol{ef}]\}}(70,1)  \label{70}
\end{equation}%
and symmetric and antisymmetric $\chi $ multiplets, $\chi _{\{i\boldsymbol{c}%
,l\boldsymbol{d}\}}$ and $\chi _{\lbrack i\boldsymbol{c},l\boldsymbol{d}]}$,
whose scalar components look as%
\begin{equation}
\chi _{\lbrack il\boldsymbol{]}[\boldsymbol{cd}]}(28,1),\text{ \ }\chi
_{\lbrack \boldsymbol{cd}]\{il\boldsymbol{\}}}(36,1)  \label{28}
\end{equation}%
Decomposing them into the components of the final $SU(5)\times SU(3)_{F}$
symmetry one finds the full set of scalars 
\begin{eqnarray}
70 &=&(5,1)+(\overline{5},1)+(10,\overline{3})+(\overline{10},3)  \notag \\
28 &=&(5,3)+(10,1)+(1,\overline{3})  \notag \\
36 &=&(5,3)+(15,1)+(1,6)  \label{36}
\end{eqnarray}%
containing the $SU(5)$ quintets $(5,1)$ and $(\overline{5},1)$ to break the
Standard Model at the electroweak scale $M_{SM}$ and the the $SU(3)_{F}$
triplet and sextet, $(1,\overline{3})$ and $(1,6)$, to properly break the
family symmetry at some large scale $M_{F}$. One may refer to the scalars (%
\ref{70}) and (\ref{28}) as the "vertical" and "horizontal" ones,
respectively, which are actually the simplest choice to form the above
Yukawa couplings. Working in pairs in them, they presumably determine masses
and mixings of all quarks and leptons. And the last but not the least, they
may be indeed composed, in the model considered, from the same preons as
quarks and leptons \cite{ch}.

\section{Conclusion}

We have investigated the potential of the local $SL(2N,C)$ symmetry to unify
all fundamental forces, including gravity. The key idea is that this
symmetry can be "trimmed down" in a controllable covariant way by the
accompanied tetrad fields. These tetrads are assumed not only determine the
spacetime geometry but also function as a kind of discerning filter,
dictating which local symmetries can operate within spacetime and how the
related elementary forces interact via gauge fields. As a result of this
filtering, the theory ends up with a simpler and more effective symmetry, $%
SL(2,C)\times SU(N)$, that translates to two separate but interacting parts:
the local $SL(2,C)$ symmetry describes gravity as a gauge force, while the
local $SU(N)$ symmetry represents a grand unified theory for other forces.

For the gravitational part, its unification with other interactions
necessitates the inclusion of quadratic curvature terms in the gravitational
sector. However, this would be experimentally unacceptable since the
coupling constant in these terms may typically be comparable to that of
vector fields. Fortunately, despite both vector and tensor fields are
located in the same $SL(2N,C)$ gauge multiplet, the tensor field
submultiplet appears naturally suppressed through the tetrad filtering that
enables the neglect of quadratic curvature terms. An essential problem
related to this type of HUT models is\ also a possible presence of ghosts
being related to the tensor rather than vector submultiplet of the $SL(2N,C)$%
\ gauge multiplet. But, again, as the quadratic strength terms of tensor
fields appears to be significantly diminished in the theory this problem is
proving to be quite surmountable. As a result, one eventually comes to the
conventional Einstein-Cartan type gravity with an extra four-fermion (spin
current-current) interaction (\ref{6}) properly suppressed in the theory.

For the grand unification, in turn, since all states involved in $SL(2N,C)$
theories are additionally classified by spin magnitude, the $SU(N)$ GUTs
with purely antisymmetric matter multiplets, including the usual $SU(5)$
theory, turn out to be irrelevant for the standard $1/2$-spin quarks and
leptons. Meanwhile, the $SU(8)$ grand unification with the certain mixed
representation for all three families of composite quarks and leptons,
arising from the $SL(16,C)$ theory, appears to be particularly interesting
and was studied in some detail. Indeed, starting from $N$ lefthanded and $N$%
\ righthanded preons being fundamental multiplets of the "metaflavor" $%
SL(2N,C)$ \ symmetry one can show, under some natural conditions, that among
all possible chiral symmetries only the $SU(8)_{L}\times SU(8)_{R}$ symmetry
meets the anomaly matching condition and can in principle ensure
masslessness of lefthanded\ and\ righthanded fermion composites at large
distances. This, in turn, identifies -- among all metaflavor $SL(2N,C)$
symmetries -- just the $SL(16,C)$ one as the most likely candidate for
hyperunification of all elementary forces.

At the same time, it is important to clarify that $SL(2N,C)$
hyperunification does not imply a single universal coupling constant for
gravity and other interactions, as is usually assumed in unified theories.
Instead, it suggests that all these forces are provided by vector and tensor
fields being the members of the same $SL(2N,C)$ gauge multiplet. A universal
constant is indeed necessary for the standard quadratic strength terms of
vector and tensor fields. However, the pure gravitational interaction has a
fundamentally different coupling, linear in the tensor field strength (\ref%
{p2}). This unique coupling arises solely due to the presence of tetrads,
which are essential ingredients for an $SL(2N,C)$ invariant theory. It comes
with its own independent coupling constant ($1/2\kappa $), conventionally
related to the Planck mass. Significantly, the vector (and axial-vector)
fields cannot have these linear strength terms alongside the standard
quadratic ones.

We should emphasize the special role of tetrads and their invertibility
condition (\ref{or7}) within the entire $SL(2N,C)$ theory framework. A key
point is that, regardless of the filtering process, tetrads must remain
truly neutral, devoid of any $SU(N)$ hyperflavored components. Otherwise,
they cannot be treated as standard vielbein fields satisfying the
invertibility conditions. This could result from spontaneous $SL(2N,C)$
breaking in the tetrad sector, as previously argued. One might also envision
that such a tetrad form might emerge at a more fundamental level if
gravitational interaction, as described by Einstein--Cartan theory, arises
as a by-product of spontaneous symmetry breaking of some primary Yang-Mills
type symmetry in a pre-geometric four-dimensional spacetime \cite{wil, wil1}%
. This would require that this gauge symmetry of the principal bundle be
properly extended to include the internal $SU(N)$ symmetry degrees as well.
In this scenario, tetrads free from this symmetry may dominate in the
spontaneously broken phase, thus leading to the gravity unification picture
discussed above.

Remarkably, only such tetrads perform the filtering process that
significantly weakens the tensor field submultiplet, while leaving the
vector and axial-vector fields unaffected. The axial-vector fields may pose
some problem in the theory, which, as argued earlier, could be resolved
through their condensation \cite{jpl} or a double filtering mechanism that
completely expels them from the theory. However, there exists an essentially
different approach that could generically solve this problem just for the
tetrads considered. Notably, unlike the vector and tensor fields,
axial-vector fields have no direct coupling with fermion matter, as was
demonstrated above. This suggests a new scenario for hyperunification,
wherein all gauge fields of the $SL(2N,C)$ symmetry appear as the composite
bosons formed by fermion pairs, rather than being elementary fields. This
approach, well-known for decades as a viable alternative to conventional
quantum electrodynamics \cite{bjorken}, gravity \cite{ph}, and Yang-Mills
theories \cite{eg,suz,jpl2}, has never been applied to noncompact unified
symmetries. In such a scenario, where only the global $SL(2N,C)$ symmetry is
initially postulated for the pure fermionic Lagrangian with appropriately
filtered fermion currents, one could expect that only composite vector and
tensor fields emerge in an effective gauge theory, while axial-vector fields
are never formed.

Another avenue for further study concerns the phenomenological aspects of
the theory. The spontaneous breaking of the $SL(2N,C)$ HUT through the
filtered $SL(2,C)\times SU(N)$ symmetry down to the Standard Model and below
will lead to many new processes caused by generalization of both gravity and
SM sectors due to new particles and new couplings, as partially discussed
above.

We may return to these important issues elsewhere.

\section*{Acknowledgments}

I would like to thank Colin Froggatt and Holger Nielsen for interesting
discussions. \bigskip

\bigskip %
%
%
%
%
%
%
%
%
%%%%%%%%%%%%%%%%%%%%%%%%%%%%%%%%%%%%%%%%%%%%%%%%%
%%%%%%
\thispagestyle{empty}

\end{document}